# Diffusive Propagation of Exciton-Polaritons through Thin Crystal Slabs

D.A. Zaitsev[1], N.D. Il'inskaya[1], A.V. Koudinov[1,2], N.K. Poletaev[1], A.Yu. Egorov[1], A.V. Kavokin[2,3] and R.P. Seisyan[1]

[1] A.F. Ioffe Physical Technical Institute, Russian Academy of Sciences, 26, Politechnicheskaya, 194021, St-Petersburg, Russia

[2] Spin Optics Laboratory, St-Petersburg State University, 1, Ulianovskaya, 198504, St-Petersburg, Russia

[3] Physics and Astronomy, University of Southampton, Highfield, Southampton, SO171BJ, United Kingdom

Comparative studies of the resonant transmission of light in the vicinity of exciton resonances, measured for 15 few-micron GaAs crystal slabs with different impurity concentrations $N$, reveal an unexpected tendency. While $N$ spans almost five decimal orders of magnitude, the normalized spectrally-integrated absorption of light scales with the impurity concentration as $N^{1/6}$. We show analytically that this dependence is a signature of the diffusive mechanism of propagation of exciton-polaritons in the semiconductor.

Exciton-polaritons are mixed light-matter quasiparticles which may be excited by light in the spectral vicinity of exciton resonances [1]. They combine properties of excitons and photons, which makes them promising for applications in opto-electronics [2]. In particular, the group velocity of an exciton-polariton may vary from the speed of light in the semiconductor crystal down to the speed of a mechanical exciton which is several orders of magnitude lower depending on the relative weight of excitonic and photonic fractions [3]. Due to their excitonic component, exciton-polaritons efficiently interact with acoustic phonons eventually decaying non-radiatively. Due to their photonic component, polaritons can decay radiatively to vacuum photonic modes outside the sample [4].

For these reasons, propagation of exciton-polaritons in crystal slabs is governed by interplay of radiative and non-radiative decay processes. When inside the crystal, a polariton can only decay non-radiatively, while once it crosses the crystal slab and achieves one of the surfaces, it decays radiatively contributing to the reflectivity or transmission signal [5]. The spectrally resolved absorption of light in a crystal $A$ is related to the reflection $R$ and transmission $T$ by the energy conservation low:

$$A(\omega) + R(\omega) + T(\omega) = const . \qquad (1)$$

Here we incorporated in the absorption all non-radiative losses including also the scattering to wave-guiding modes which decay through the side edges of the sample. It is hard to measure the absorption directly, but one can deduce $A(\omega)$ from the transmission spectra as

$$A(\omega) \sim -\ln T(\omega), \qquad (2)$$

if one assumes an exponential decay of the light intensity through the crystal. This assumption is generally valid in semiconductor films where the Perot-Fabry interference is suppressed because of the absorption, and the modulation of reflectivity in the vicinity of the exciton resonance does not exceed 3% [6]. More accurate extraction of the true absorption spectrum from the transmission spectrum can be done, e.g., using the transfer matrix method [7].

In this Letter, we present the experimental study of the spectrally-integrated excitonic absorption of light $K$ measured for 15 similar GaAs crystal slabs which differ by the concentrations of impurity centers $N$. While the excitonic absorption is an intrinsic effect and, at a first glance, should not depend on $N$ at all, we have found a surprising sublinear dependence $K \propto N^\beta$ with $\beta$ close to $1/6$. This observation sheds light on the mechanisms of non-radiative losses in semiconductors and allows concluding on the diffusive propagation of exciton-polaritons in our samples.

The samples for our studies were based on the epitaxial layers of GaAs, grown on the bulk semi-insulating GaAs substrates by means of either molecular beam epitaxy (MBE) or vapor phase epitaxy (VPE). The epitaxial layers were nominally *p*-doped during the growth process. The impurity concentrations were then probed by either voltage–capacitance (VC) or Hall-effect measurements at room temperature. The measured values of the impurity concentration $N$ are listed in Table 1 together with the nominal dopants and other relevant information.

After that, the substrates were removed by etching in ammonium and hydrogen peroxide water solution, and the epitaxial layers were etched down to few-micrometer thicknesses. The resulting thin crystal slabs were annealed in the atmosphere of hydrogen (in order to remove tension and oxides remaining on the surface after etching) and then loosely packed between two cover-glasses and sealed in the air. During optical measurements, these sandwiches were immersed in liquid helium at 2 K.

As a first step, we measured the photoluminescence (PL) spectra for a selection of our samples in order to check possible effects of previous etching and sandwiching on the actual concentration of the impurity states. The PL was excited by the second harmonic of continuous-wave YAG:Nd laser (2.33 eV); the spectra were taken using a diffraction spectrometer and a photomultiplier. All in all, we have found that our slabs produce quite typical PL spectra of GaAs at the corresponding levels of doping. The observed features of the PL spectra can be readily assigned to various intrinsic or impurity-related optical transitions known from the optical spectra of bulk GaAs, as shown in Fig.1. Remarkably, the PL spectrum of the slab with nominal $N = 1 \cdot 10^{13}$ cm$^{-3}$ shows a single line of the free-exciton transition; the impurity-related features are not seen. This observation, first, confirms a very high purity of the F235 sample and, second, suggests that previous treatment did not really change the impurity concentration in our slabs. At the same time, the presence of the acceptor-related and donor-related lines simultaneously in one spectrum, as observed for some specimens (see upper two spectra in Fig.1), points to a degree of compensation.

As a second step, we measured the transmission spectra of the samples. The "white light" from the incandescent lamp was first passed through the red filter which cut off the high-energy photons (>1.9 eV), then focused on and passed through the sample, and finally focused on the entrance slit of a diffraction spectrometer. The pump density was about 1 W/cm$^2$. The raw data on the spectrally-resolved transmission $I(\omega)$ for every sample were then normalized by the spectrum of the lamp. To this end, we had taken the reference spectrum $I_0(\omega)$ of the lamp "white light" transmitted through the sample-box containing no sample inside.

At the next stage, the transmission data were processed further. We found the logarithm of the normalized transmission $I(\omega)/I_0(\omega)$ to obtain, having reversed the sign, the spectrum of optical density $A(\omega) = \alpha(\omega)d$. We aim at extracting the absorption coefficient $\alpha(\omega)$ which should be an immanent characteristic of a corresponding semiconductor; however, the direct calculation is hindered because the slab thickness $d$ is not accurately known and, moreover, can be not quite homogeneous widthways. Thus we took a straightforward assumption that the high-energy (interband) absorption coefficient should have the same value for any doping level, and we scaled the experimental optical density spectra of every sample in such a way that the high-energy shelves match the textbook value, about 8000 cm$^{-1}$ for the bulk GaAs [8] (see Fig.2).

The absorption spectra shown in Fig.2 demonstrate a distinct excitonic line peaked slightly above 1.515 eV, in the vicinity of the fundamental absorption edge for the interband transitions. While the peak energy is similar for all the samples, spectral widths and heights of the excitonic line vary for different samples. Thus, an integral value like the area under the excitonic absorption contour $K = \int A(\omega)d\omega$ should be the most adequate characteristic of the efficiency of the excitonic absorption in a particular sample. One can clearly see the enhancement of the integrated exciton absorption with the increasing impurity concentration $N$ (from lower to higher spectra in Fig.2).

For a quantitative analysis, the experimental spectra in Fig.2 should be decomposed into the excitonic and non-excitonic contributions. The latter can include not only the classical fundamental absorption edge (whose spectral dependence might be known well enough), but the Urbach-like density-of-states tails as well, e.g., due to the Franz-Keldysh effect or disorder effects. The methods of analyses of the shape of the exciton absorption peak are described elsewhere [9,10]. Here we focus on the unusual dependence of the integrated absorption $K$ on the impurity concentration $N$.

We assumed that the low-energy part of every spectrum in Fig.2, up to the energy of the excitonic maximum, was not affected by the non-excitonic absorption. We then mirror-reflected this part to the right and calculated the area under the resulting bell-shaped contour. The obtained values of $K$, the spectrally-integrated excitonic absorption, are plotted in Fig.3 against the corresponding impurity concentrations $N$.

Fig.3 summarizes the data on the variation of the integrated absorption with the impurity concentration. In the set of studied samples, $N$ varies in wide limits, from $N = 1 \cdot 10^{13}$ cm$^{-3}$ to about $N = 5 \cdot 10^{17}$ cm$^{-3}$. For instance, in terms of electro-physical properties, such a range spans from semi-insulating to semi-metallic behaviour of bulk GaAs. But throughout the range, the dependence $K(N)$ obeys a single power law, which is expressed by following a single straight line in a double logarithmic scale. A purely phenomenological two-parametric fit of the experimental trend by the function

$$K = \left( N / N_0 \right)^{\beta} \tag{3}$$

resulted in $\beta = 0.175 \pm 0.005$, $N_0 = 10^{8.3 \pm 0.9}$ cm$^{-3}$. We shall see below that simple theoretical considerations give the value of the factor $\beta = 1/6 \approx 0.167$.

The observed dependence of the integrated absorption on the impurity concentration is somewhat unexpected. Indeed, any kind of "impurity absorption" would be most naturally expected to show a linear dependence on $N$. Similar dependence on $N$ was predicted for exciton-polaritons by Akhmediev [11] and reproduced in the recent theoretical work [12]. On the other hand, the excitonic absorption is an intrinsic phenomenon which might have been independent on *N* at all. These naïve expectations implicitly rely on the ideas of a ballistic propagation of photons through a matter where they rarely experience acts of inelastic scattering thus transferring photon energy to the crystal lattice. In contrast to that, in the presence of strong elastic scattering, the effective trajectory of each exciton-polariton propagating through the crystal may get significantly longer, which leads the "slow light" phenomenon [13]. Below we show that in this regime, which we shall refer to as the diffusive propagation regime, the integrated absorption is governed by the characteristic time spent by diffusing polaritons in the crystal slab, which in turn depends non-linearly on the impurity concentration.

The propagation of exciton-polaritons in a semiconductor slab containing a sufficiently high concentration of the diffusive centers can be described by a classical diffusion equation:

$$\frac{\partial n(\omega,x,t)}{\partial t} = D(\omega)\frac{\partial^2 n(\omega,x,t)}{\partial x^2} - \frac{n(\omega,x,t)}{\tau(\omega)}, \qquad (4)$$

where $n(\omega,x,t)$ is the polariton density, $D(\omega)$ is the frequency dependent diffusion coefficient, $\tau(\omega)$ is the exciton-polariton non-radiate life-time, *x* is the coordinate in the direction normal to the plane of the slab.

The diffusive propagation time of an exciton-polariton through the crystal slab of the thickness *L* can be found from Eq. (4) according to Ref. [13]

$$T_D(\omega) = \sqrt{\left(\frac{\tau(\omega)}{4}\right)^2 + \frac{L^2\tau(\omega)}{4D(\omega)}} - \frac{\tau(\omega)}{4}. \qquad (5)$$

If the polariton non-radiative life-time is short enough (the strong absorption case) so that

$$\tau(\omega) \ll \frac{L^2}{D(\omega)}, \qquad (6)$$

Eq. (5) reduces to

$$T_D(\omega) = \frac{L}{2}\sqrt{\frac{\tau(\omega)}{D(\omega)}}. \qquad (7)$$

The diffusion coefficient is dependent on the mean free path of the exciton polariton $l(\omega)$ as

$$D(\omega) = \frac{1}{3}l(\omega)v_{gr}(\omega). \qquad (8)$$

If the coherence length of exciton-polaritons strongly exceeds the mean spacing between impurities, the polariton mean free path scales with the concentration of scattering centers as

$$l(\omega) \sim N^{-1/3}. \qquad (9)$$

This regime, which contrasts with the regime of diffusion by short range interactions in a classical gas, is characteristic for exciton-polaritons which are macroscopic quasiparticles with a coherence length frequently amounting to several μm [14], while the average distance between impurity centers is of the order or less than 10 nm (see Table 1). The proportionality (9) can be derived in many ways, but the simplest argument proving its validity is the dimensionality argument. If the coherence length is much larger than any characteristic length of the scattering problem, it can be assumed infinite. The relation between two remaining characteristic lengths which are the mean free path and the mean distance between impurities cannot be anything but proportionality in this case, that readily yields the functional dependence (9).

Substitution of (9) into Eq. (7) yields

$$T_D(\omega) \sim N^{1/6}. \qquad (10)$$

Now, the integrated absorption can be estimated as

$$K \sim \int \frac{T_D(\omega)}{\tau(\omega)} d\omega \sim N^{1/6}. \qquad (11)$$

One can see that the diffusion model accurately reproduces the experimentally found functional dependence of the integrated absorption on the concentration of impurity centers in a semiconductor. This shows that, in the presence of impurities, propagation of exciton-polaritons has a diffusive character. We note here that the strong absorption limit applies only to relatively slow exciton-polaritons which are characterized by a high exciton component. These polaritons spend longer time inside the crystal slab and have better chances to decay non-radiatively than fast photon-like polaritons.

In conclusion, the proof of diffusive propagation of exciton polaritons in semiconductor slabs has been obtained by a systematic study of the integrated absorption in a series of 15 GaAs samples. The dependence of the integrated absorption on the concentration of impurity centers unambiguously shows that the absorption of light is mostly through slow exciton-like exciton-polaritons which experience multiple scattering acts while propagating across the slab. This observation is important for understanding of mechanisms of the energy transfer between light and matter in semiconductors.

*Acknowledgement.* AK and AK acknowledge the support from the Russian Ministry of Science and Education (contract no. 11.G34.31.0067). This work was partially supported by RFBR (project 13-02-00316). The experimental part of the work was supported by Committee on Science and Higher Education of the Government of St. Petersburg.

**Table 1.** Parameters of the samples used in this study.

| Sample Label | Growth Method | Impurity concentration, N, cm$^{-3}$ | N measured using |
|---|---|---|---|
| F235-A,B | VPE | $1 \cdot 10^{13}$ | VC |
| R1488-1,3,4 | MBE | $2 \cdot 10^{14}$ | VC |
| F262-A | VPE | $8 \cdot 10^{14}$ | VC |
| F244-C | VPE | $1.3 \cdot 10^{15}$ | Hall&VC |
| F247-A,B,D | VPE | $8 \cdot 10^{15}$ | Hall |
| R1278-1,2 | MBE | $2.8 \cdot 10^{16}$ (Si) | Hall |
| R1144-2 | MBE | $1.85 \cdot 10^{17}$ (Si) | Hall |
| R1275-1,3 | MBE | $4.8 \cdot 10^{17}$ (Si) | Hall |

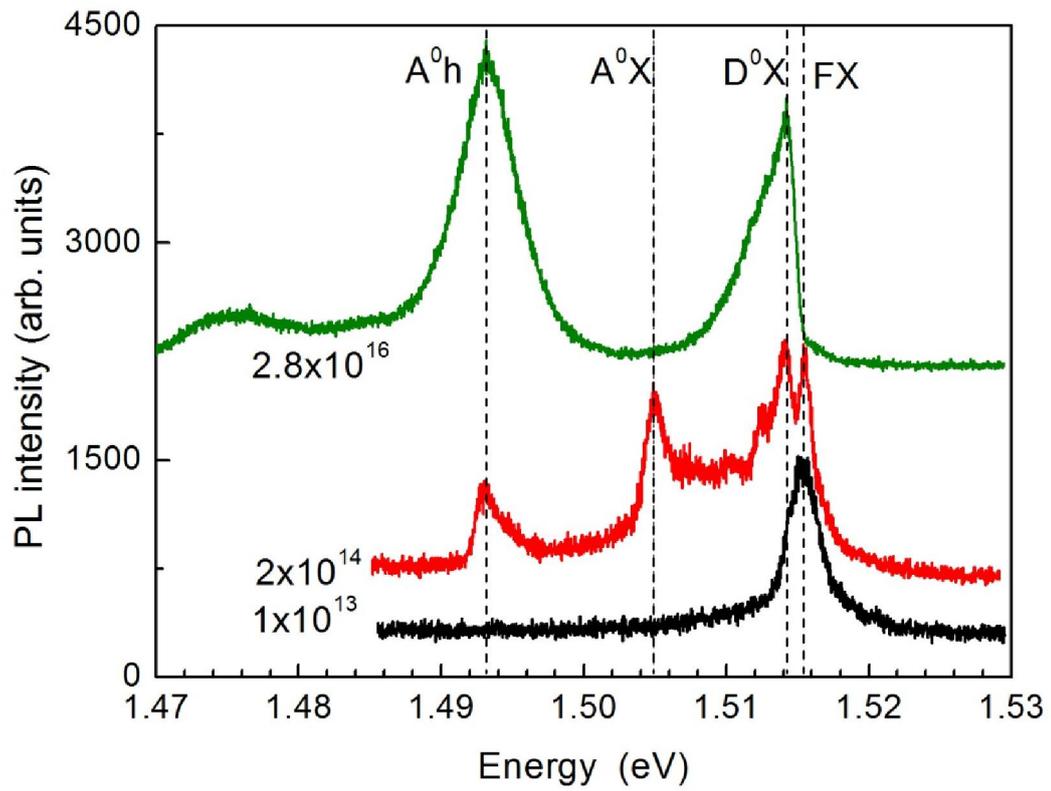

**Figure 1.** Representative PL spectra of sandwiched GaAs slabs: samples F235-B, R1488-1 and R1278-2. Excitation at 2.33 eV, T=2K. Assignment of the peaks as labeled.

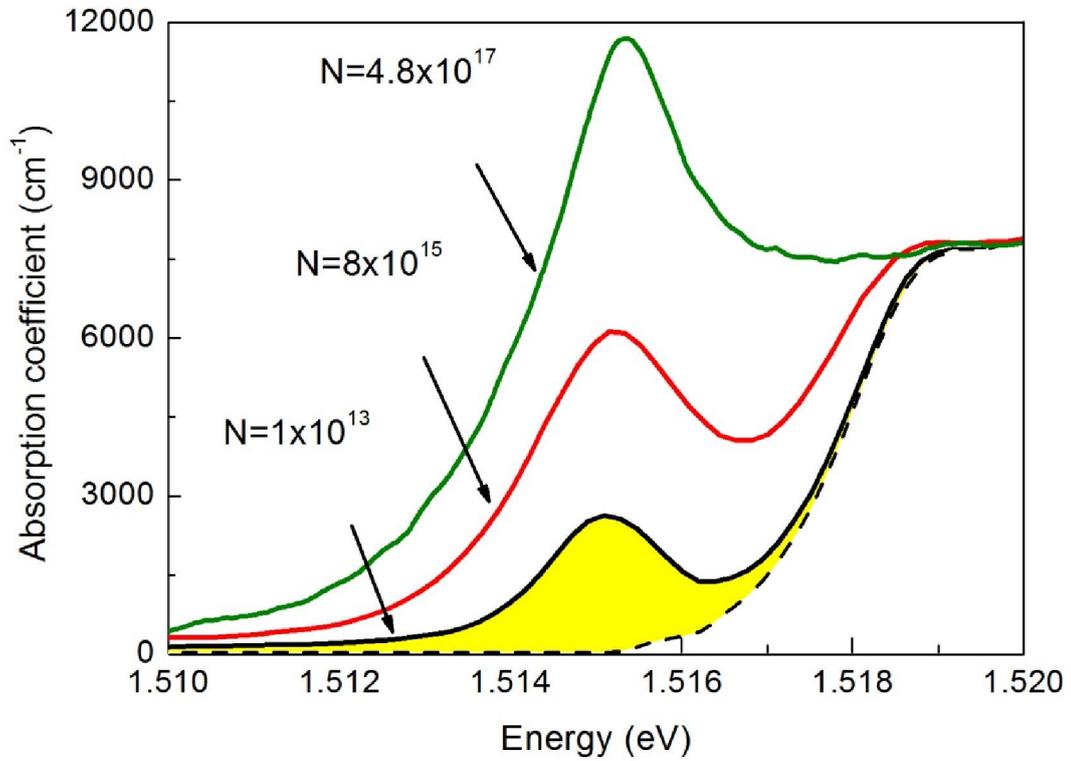

**Figure 2.** Representative spectra of the absorption coefficient in the vicinity of the fundamental absorption edge: samples F235-A, F-247-B, R1275-1. The spectra (solid lines) were extracted from the light transmission data as described in the text. Decomposition of the lowest spectrum into the excitonic contour (shaded area) and the background of the fundamental absorption (dashed line) is shown.

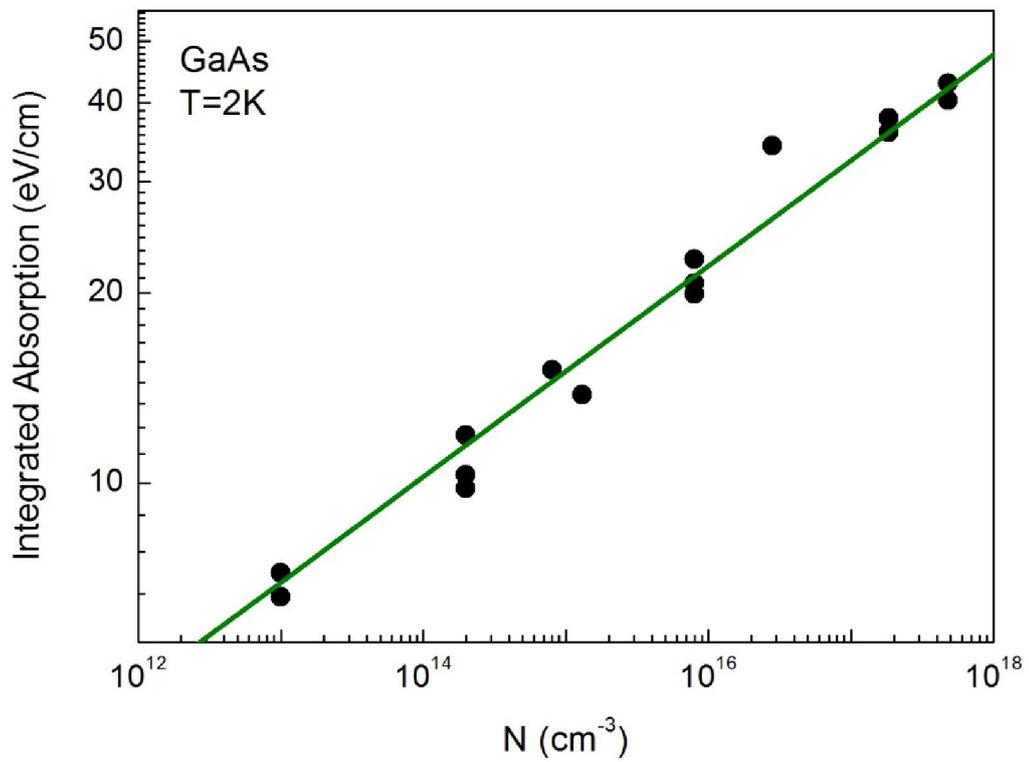

**Figure 3.** Energy-integrated excitonic absorption *K* versus the impurity concentration. Each experimental point represents a separate sample (see Table 1). Straight line shows the prediction of a diffusive model formulated here (Eq. (11)).